\documentclass[twocolumn]{aastex701}

\usepackage{fontawesome}
\usepackage{amsmath}
\defcitealias{2026ApJ..1006...61W}{Paper~I}

\begin{document}

\title[Predicting Concentration for Individual Dark Matter Halo]{Predicting Individual Dark Matter Halo Concentrations from Their Assembly Histories}

\correspondingauthor{Weipeng Lin\\ \hspace*{\parindent}\href{mailto:linweip5@mail.sysu.edu.cn}{linweip5@mail.sysu.edu.cn}}
\correspondingauthor{Dao-Zhou Wang\\ \hspace*{\parindent}\href{mailto:wangdzh25@mail.sysu.edu.cn}{wangdzh25@mail.sysu.edu.cn}}

\author[orcid=0000-0002-5358-3305,sname='Wang']{Dao-Zhou Wang}
\affiliation{School of Physics and Astronomy, Sun Yat-sen University, DaXue Road 2, 519082 Zhuhai, China}
\email{wangdzh25@mail.sysu.edu.cn}

\author[orcid=0000-0003-2204-2474, sname='Lin']{Weipeng Lin} 
\affiliation{School of Physics and Astronomy, Sun Yat-sen University, DaXue Road 2, 519082 Zhuhai, China}
\affiliation{CSST Science Center for the Guangdong-Hongkong-Macau Greater Bay Area, DaXue Road 2, 519082 Zhuhai, China}
\email{linweip5@mail.sysu.edu.cn}

\author[orcid=0000-0002-1189-2855, sname='Zhu']{Weishan Zhu} 
\affiliation{School of Physics and Astronomy, Sun Yat-sen University, DaXue Road 2, 519082 Zhuhai, China}
\affiliation{CSST Science Center for the Guangdong-Hongkong-Macau Greater Bay Area, DaXue Road 2, 519082 Zhuhai, China}
\email{zhuwshan5@mail.sysu.edu.cn}

\author[orcid=0009-0001-6592-6467,sname='Luan']{Tian-Cheng Luan} 
\affiliation{School of Physics and Astronomy, Sun Yat-sen University, DaXue Road 2, 519082 Zhuhai, China}
\email{dingdluan@outlook.com}

\begin{abstract}

We present an empirical model for predicting concentration parameters of individual dark matter halos in cold and warm dark matter (WDM) cosmologies over \(z=0\)–5. Constructed using a large suite of cosmological N-body simulations, our model separately incorporates cosmic expansion (linear growth factor), mass scale (linear density fluctuations), halo mass assembly histories via the Extended Press–Schechter (EPS) scaled formation variable, plus a WDM damping term tied to the half-mode mass. It works for both virial and \(200\rho_\mathrm{crit}\) halo definitions. Across all tested cosmologies and redshifts, our global logarithmic fit gives concentration scatter \(\sim0.09\) dex for relaxed halos and \(\sim0.09\)–\(0.10\) dex for all halos. While relaxed halos maintain nearly constant scatter near 0.09 dex, the scatter of all halos at \(z>3\) increases from 0.10 to 0.18 dex, revealing substantial violent-relaxation impacts on high-redshift halo structure. At fixed mass, concentration scatter is dominated by assembly history. We extract the scaled formation variable’s probability distribution function (PDF) from merger trees and find departures from standard EPS theory. This PDF is narrower in WDM than \(\Lambda\)CDM due to suppressed small-scale perturbations and rapid, delayed monolithic collapse near the WDM truncation scale. Monte Carlo realizations with our formulae reproduce concentration distributions and match our prior peak-concentration model, whereas standard EPS underestimates concentrations from delayed-formation bias. Our framework offers a transparent, computationally lightweight tool to predict individual halo concentrations across diverse cosmological models.

\end{abstract}


\section{Introduction} \label{sec:intro}

The concentration parameter is a fundamental descriptor of the internal structure of dark matter halos. Within the standard NFW \citep{1996ApJ...462..563N,1997ApJ...490..493N} description of halo density profiles, it characterizes the relative size of the halo boundary and the inner scale radius, and therefore provides a compact measure of how centrally condensed a halo is. Concentration is widely used in studies of nonlinear structure formation, the galaxy--halo connection, gravitational lensing, and dark matter phenomenology.

A large number of studies have established that halo concentration depends systematically on halo mass, redshift, and cosmology (e.g., \citealp{2000ApJ...535...30J}; \citealp{2001MNRAS.321..559B}; \citealp{2001ApJ...554..114E}; \citealp{2002ApJ...568...52W}; \citealp{2003ApJ...597L...9Z}; \citealp{2005MNRAS.357...82R};  \citealp{2007MNRAS.378...55M}; \citealp{2007MNRAS.381.1450N}; \citealp{2008MNRAS.387..536G}; \citealp{2008MNRAS.390L..64D}; 
\citealp{2009ApJ...707..354Z}; 
\citealp{2012MNRAS.427.1322L, 2014MNRAS.441..378L}; \citealp{2014MNRAS.441.3359D};  \citealp{2015ApJ...799..108D}; \citealp{2019ApJ...871..168D}; \citealp{2021MNRAS.506.4210I}; \citealp{2024MNRAS.52710760W}; \citealp{2026ApJ..1006...61W}; hereafter \citetalias{2026ApJ..1006...61W}). Most of this work, however, has focused on population-level statistics, such as the mean or median concentration--mass relation. In our previous work (\citetalias{2026ApJ..1006...61W}), 
we argued that the peak of the concentration distribution provides a robust characterization of the most typical halo population. 
By extending the conventional peak-height framework to include information from halo mass accretion histories (MAH), we constructed a formation peak-height parameter and used it to derive universal fitting formulae for the peak concentration over a broad range of cosmologies and redshifts.

A complementary question is how to describe the halo-to-halo scatter at fixed mass and redshift. This issue is physically important because halos with the same virial mass observed at the same epoch can still show substantial structural diversity. 
Such scatter directly affects applications that depend on halo internal structure. In the studies of galaxy formation and evolution, concentration governs the central gravitational potential well and inner circular velocity profile, thereby modulating gas cooling, galaxy sizes, rotation curves, and assembly bias in the galaxy--halo connection. In large-scale structure analyses, capturing halo-to-halo structural variations is essential for modeling small-scale nonlinear clustering, predicting gravitational lensing signals, and constructing realistic mock catalogs for wide-field galaxy surveys.

A long-standing idea is that halo-to-halo scatter in concentration is closely tied to mass accretion history. Earlier studies showed that earlier-forming halos are generally more concentrated (e.g., \citealp{2002ApJ...568...52W}; \citealp{2003MNRAS.339...12Z};
\citealp{2009ApJ...707..354Z}), indicating that concentration retains memory of halo assembly. 
\citet{2016MNRAS.460.1214L} developed a physically motivated model that links the characteristic inner density to the critical density at the epoch when the corresponding mass scale first collapsed. This model reproduces the median concentration–mass relation and, given the complete collapse‑mass history of all progenitors, also predicts individual‑halo concentrations in both $\Lambda$CDM and $\Lambda$WDM cosmologies. 
More recently, \citet{2025Univ...11...37Z} showed with neural networks that MAH information substantially improves halo-by-halo concentration predictions, although their reported root mean square error (RMSE) in logarithmic concentration space increases from 0.0845 to 0.1287 over $z=0$--2. 
However, systematic tests and calibrations of individual halo concentration predictions beyond $z = 2$ remain scarce.
Nevertheless, these studies suggest that concentration scatter largely reflects the diversity of halo formation histories, while leaving open whether individual concentrations can be captured by a compact, physically interpretable, and nearly universal relation across redshift, cosmology, and halo definition.

This work aims to develop a simple and compact empirical model for the concentration parameter of individual halos. While \citetalias{2026ApJ..1006...61W} encapsulated the mass and redshift dependence of the population peak concentration into an effective formation peak height, in this work we decouple the cosmological linear growth factor $D(z)$ and the linear density rms fluctuation $\sigma(M)$ as independent regressors. This decoupled formulation provides enhanced flexibility to independently capture cosmic expansion background effects and initial power spectrum amplitude scalings. To parameterize the individual MAH of each simulated halo, we utilize the EPS-scaled formation variable \(\omega_{\rm f}\) \cite[][see below for definition]{1993MNRAS.262..627L}. 
For WDM cosmologies, we further include the half-mode mass $M_{\rm hm}$, which characterizes the cutoff scale in the linear power spectrum and is defined in detail in Section~\ref{sec:variables}. 
Using a broad suite of cosmological simulations, we primarily calibrate an empirical, compact relation for dynamically relaxed halos that connects individual halo concentrations to their collapse rarity, assembly history, and the WDM cutoff scale. We also provide the corresponding relation for the complete all-halo sample for comparison. We first verify this formulation through a direct halo-by-halo comparison with simulation measurements across redshifts $z=0$--5. 
We then demonstrate how the overall concentration distribution can be statistically reconstructed from the halo formation-time distribution via Monte Carlo realizations, showing excellent consistency with the peak-concentration relations established in \citetalias{2026ApJ..1006...61W}. Finally, we compare our empirical framework against existing physically motivated and machine-learning models to highlight its robustness, computational efficiency, and physical interpretability across diverse cosmological scenarios.

This paper is organized as follows. Section~\ref{sec:data} describes the simulation samples, relaxed halo selection, and model variables. Section~\ref{sec:results} presents the empirical relation, tests it with EPS- and simulation-based formation times, and compares the predicted concentrations with simulations and previous individual halo models. Section~\ref{sec:discussion} summarizes our conclusions. Throughout this paper, `log' denotes a base-10 logarithm.

\section{Data and Method}
\label{sec:data}

\subsection{Numerical simulations}
\label{sec:simulations}

The simulations used in this work are based on the data set of \citetalias{2026ApJ..1006...61W}, supplemented by the Uchuu simulation suite. We use the dark-matter-only runs from the TNG project\footnote{\url{https://www.tng-project.org}} (\citealp{2018MNRAS.480.5113M}; \citealp{2018MNRAS.477.1206N}; \citealp{2018MNRAS.475..624N}; \citealp{2018MNRAS.475..648P}; \citealp{2018MNRAS.475..676S}), the MultiDark-Planck simulations\footnote{\url{https://www.cosmosim.org}} \citep{2016MNRAS.457.4340K}, and the Uchuu simulations\footnote{\url{http://skiesanduniverses.org}} \citep{2021MNRAS.506.4210I} as our $\Lambda$CDM samples. These public simulations provide a broad coverage in halo mass and numerical resolution, from low-mass halos resolved in small high-resolution boxes to massive cluster-scale halos sampled in large cosmological volumes.

To test the dependence on cosmology, we also include the non-$\Lambda$CDM simulations described in \citetalias{2026ApJ..1006...61W}. 
These simulations were performed specifically for our project
and consist of an SCDM model, an OCDM model, a scale-free Einstein--de Sitter (EdS) model with an initial power spectrum $P(k)\propto k^{-2}$, two $w$CDM models, and a set of $\Lambda$WDM models. The two $w$CDM simulations adopt constant dark-energy equation-of-state parameters corresponding to the first and 37th models of the Coyote suite (see \citealp{2009ApJ...705..156H}, \citealp{2010ApJ...715..104H} for details), while the $\Lambda$WDM simulations use equivalent thermal relic particle masses of $m_{\rm WDM}=0.5$, $0.8$, $1.1$, and $1.5\,{\rm keV}$.

The main numerical settings and cosmological parameters of all simulations are summarized in Table~\ref{tab:simulations}. The full sample spans box sizes of $30$--$2500\,h^{-1}{\rm Mpc}$ and particle masses of $3.65\times10^5$--$2.36\times10^{10}\,h^{-1}\mathrm{M_\odot}$, allowing us to study halo concentrations over a wide range of masses, redshifts, resolutions, and cosmological models.

\subsection{Halo catalogs and concentration measurements}
\label{sec:halo_catalogs}

Halo catalogs and merger trees are taken from public releases whenever available. 
For the $\Lambda$CDM cosmological simulations,
we use the corresponding public halo catalogs and merger-tree products. For the simulations generated in this project, halos are identified with the phase-space halo finder Rockstar (\citealp{2013ApJ...762..109B}), and merger trees are constructed with Consistent trees (\citealp{2013ApJ...763...18B}).

Unless otherwise stated, our main analysis adopts the virial halo definition, for which the overdensity $\Delta_{\rm vir}$ is computed using the fitting formula of \citet{1998ApJ...495...80B} for each cosmology and redshift. More generally, for a halo defined by an overdensity threshold $\Delta$, we denote its mass and radius by $M_{\Delta}$ and $r_{\Delta}$, and define the concentration parameter as
\begin{equation}
    c_{\Delta} \equiv \frac{r_{\Delta}}{r_{\rm s}},
\end{equation}
where $r_{\rm s}$ is the NFW scale radius, often denoted by $r_{-2}$, namely the radius at which the logarithmic slope of the density profile equals $-2$. Our fiducial results use $\Delta=\Delta_{\rm vir}$ and are denoted by $M_{\rm vir}$, $r_{\rm vir}$, and $c_{\rm vir}$. We also present results for $\Delta=200c$, denoted by $M_{200c}$, $r_{200c}$, and $c_{200c}$, in order to compare with previous work and to quantify the dependence of our concentration relations on the halo overdensity definition.

For the MultiDark-Planck, Uchuu, and our own simulations, halo concentrations are taken from the corresponding halo catalogs. The TNG public catalogs provide only $c_{200c}$, whereas both $c_{\rm vir}$ and $c_{200c}$ are required in this work. We therefore derive the TNG concentrations by fitting NFW profiles directly, following the same procedure as in \citetalias{2026ApJ..1006...61W}. This gives a consistent set of
$c_{\rm vir}$ and $c_{200c}$ measurements for the TNG halos.

In the profile fitting, spherically averaged density profiles are constructed with 50 logarithmically spaced radial bins. For most simulations, the fits are performed over the radial range from $3\epsilon$ to $r_{\rm vir}$, where $\epsilon$ is the Plummer-equivalent softening length. As shown in \citetalias{2026ApJ..1006...61W}, $3\epsilon$ is larger than or comparable to the convergence radius $r_{\rm conv}$ (\citealp{2019MNRAS.488.3663L}) for these simulations. For the TNG simulations, however, $3\epsilon<r_{\rm conv}$ for a non-negligible fraction of halos, and we therefore set the inner fitting radius to $r_{\rm conv}$. This conservative choice avoids the unresolved inner region and yields consistent NFW concentration estimates across the simulation set.

To reduce numerical-resolution effects and ensure robust concentration measurements, we retain only halos resolved by at least 2000 dark matter particles within $r_{\Delta}$,
consistent with the conservative criteria of \citet{2018ApJ...859...55C} and \citetalias{2026ApJ..1006...61W}. Because the simulations have different mass resolutions, the corresponding minimum halo mass varies from one simulation to another; overlaps between simulations are used to check the consistency of the resulting concentration measurements.

In addition to the particle-number cut, we remove halos with unreliable merger histories or poorly resolved profile fits. We discard halos whose main progenitors were identified as satellite subhalos at any earlier snapshot, since these objects are often associated with flyby or backsplash events and can have artificially high concentrations, especially at low masses. This cut follows the spirit of \citet{2022MNRAS.509.3441A}. We also exclude low-mass halos for which the fitted NFW scale radius is comparable to the gravitational softening length, as their inner profiles are not sufficiently resolved and their fitted concentrations can be spuriously large. These cuts are applied before deriving the concentration relations below.

\begin{table*}
\centering
\small 
\setlength{\tabcolsep}{0.5pt}
\renewcommand\arraystretch{1.0}
\caption{The cosmological parameters and setting of simulations. The columns list the simulation identifier, comoving box size $L_{\rm box}$, number of dark matter particles $N_{\rm DM}$, particle mass $m_{\rm DM}$, Plummer-equivalent gravitational softening length $\epsilon$, the cosmological parameters $\Omega_m$, $\Omega_{\Lambda}$, $\sigma_8$, $n_s$, and $h$, the additional model parameter $w$ for $w$CDM cosmologies or $m_{\rm WDM}$ for WDM cosmologies, and the simulation code. }
\begin{tabular*}{\textwidth}{@{\extracolsep{\fill}} l c c c c c c c c c c c } 
\hline\hline
Simulation & $L_{\text {box }}$ &$N_{\text {DM }}$ & $m_{\text {DM }}$ & $\epsilon$ & $\Omega_{m}$ & $\Omega_{\Lambda}$ & $\sigma_{8}$ & $n_{s}$ & $h$ & $w$ or $m_\mathrm{WDM}$ & Code \\
 &$\left(h^{-1}\mathrm{Mpc}\right)$ & 
&$\left(h^{-1} M_{\odot}\right)$ &$\left(h^{-1}\mathrm{kpc}\right)$ & & & & & & (keV)\\
\hline TNG50-dark & $35$  &  $2160^{3}$  &3.65$\times10^{5}$ &0.20 &0.3089  &0.6911 &0.8159 & 0.9667 & 0.6774 &$\cdots$  & \texttt{AREPO}\\
 TNG100-dark &  $75$ & $1820^{3}$ &6.0$\times10^{6}$ &0.50 &0.3089 &0.6911 &0.8159 & 0.9667 & 0.6774 &$\cdots$  & \texttt{AREPO}\\
 TNG300-dark  & $205$ &  $2500^{3}$  &4.73$\times10^{7}$ &1.0 &0.3089 &0.6911 &0.8159 & 0.9667 & 0.6774 &$\cdots$  & \texttt{AREPO}\\
 VSMDPL &160 &$3840^{3}$  &6.20$\times10^{6}$ &2.0  &0.307 &0.693 &0.8228 & 0.960 & 0.6777 &$\cdots$  & \texttt{GADGET-2}\\
 SMDPL &400 &$3840^{3}$   &9.63$\times10^{7}$ &1.50  &0.307 &0.693 &0.8228 & 0.960 & 0.6777 &$\cdots$ & \texttt{GADGET-2}\\
 MDPL2 &1000 &$3840^{3}$  &1.51$\times10^{9}$ &5.0  &0.307 &0.693 &0.8228 & 0.960 & 0.6777 &$\cdots$  & \texttt{GADGET-2}\\
 BigMDPL &2500 &$3840^{3}$ &2.36$\times10^{10}$ &10.0  &0.307 &0.693 &0.8228 & 0.960 & 0.6777 &$\cdots$  & \texttt{GADGET-2}\\
 Micro-Uchuu & $100$  &  $640^{3}$  &3.27$\times10^{8}$ &4.27 &0.3089  &0.6911 &0.8159 & 0.9667 & 0.6774 &$\cdots$  & \texttt{GreeM}\\
 Shin-Uchuu &  $140$ & $6400^{3}$ &8.97$\times10^{5}$ &0.40 &0.3089 &0.6911 &0.8159 & 0.9667 & 0.6774 &$\cdots$  & \texttt{GreeM}\\
 Mini-Uchuu  & $400$ &  $2560^{3}$  &3.27$\times10^{8}$ &4.27 &0.3089 &0.6911 &0.8159 & 0.9667 & 0.6774 &$\cdots$  & \texttt{GreeM}\\
 SCDM &100 &$512^{3}$ &2.07$\times10^{9}$ &4.88 &1.0 &0 &0.55 & 1.0 & 0.50 &$\cdots$  & \texttt{GADGET-4}\\
 OCDM &100 &$512^{3}$ &6.20$\times10^{8}$ &4.88 &0.30 &0 &1.0 & 1.0 & 0.667 &$\cdots$  & \texttt{GADGET-4}\\
 SF-n2 &100 &$512^{3}$ &2.07$\times10^{9}$ &4.88 &1.0 &0 &1.0 & -2.0 &$\cdots$  &$\cdots$  & \texttt{GADGET-4}\\
$\Lambda$WDM-L100-0.5 & 100 & $512^{3}$ & 6.43$\times10^{8}$ & 4.88 & 0.31 & 0.69 & 1.0 & 1.0 & 0.6766 & 0.5 & \texttt{GADGET-4}\\
$\Lambda$WDM-L100-0.8 & 100 & $512^{3}$ & 6.43$\times10^{8}$ & 4.88 & 0.31 & 0.69 & 1.0 & 1.0 & 0.6766 & 0.8 & \texttt{GADGET-4}\\
$\Lambda$WDM-L100-1.1 & 100 & $512^{3}$ & 6.43$\times10^{8}$ & 4.88 & 0.31 & 0.69 & 1.0 & 1.0 & 0.6766 & 1.1 & \texttt{GADGET-4}\\
$\Lambda$WDM-L100-1.5 & 100 & $512^{3}$ & 6.43$\times10^{8}$ & 4.88 & 0.31 & 0.69 & 1.0 & 1.0 & 0.6766 & 1.5 & \texttt{GADGET-4}\\
$\Lambda$WDM-L60-1.5 & 60 & $512^{3}$ & 1.39$\times10^{8}$ & 2.93 & 0.31 & 0.69 & 1.0 & 1.0 & 0.6766 & 1.5 & \texttt{GADGET-4}\\
$\Lambda$WDM-L30-1.5 & 30 & $1024^{3}$ & 2.17$\times10^{6}$ & 0.73 & 0.31 & 0.69 & 1.0 & 1.0 & 0.6766 & 1.5 & \texttt{GADGET-4}\\
 $w$CDM1 &100 &$512^{3}$ &8.91$\times10^{8}$ &4.88 &0.4308 &0.5692 &0.8161 & 0.9468 & 0.5977 &-0.816  & \texttt{SWIFT}\\
 $w$CDM37 &100 &$512^{3}$ &5.79$\times10^{8}$ &4.88 &0.28 &0.72 &0.9000 & 1.0233 & 0.7313 &-1.294  & \texttt{SWIFT}\\
\hline
\end{tabular*}
\vspace{1.5ex}
\label{tab:simulations}
\end{table*}

\subsection{Relaxed halos}
\label{sec:relaxed_halo}

Because concentration is sensitive to the detailed shape of the halo density profile, we restrict the analysis primarily to relaxed halos. This helps remove systems that are strongly affected by recent mergers or other transient disturbances.

A halo is classified as relaxed if it satisfies
\begin{equation}
    x_{\rm off} < 0.07,\qquad \lambda < 0.07,\qquad 2T/|U| < 1.35,
\end{equation}
where $x_{\rm off}$ is the offset between the halo center and center of mass, $\lambda$ is the spin parameter, and $T$ and $U$ are the kinetic and potential energies. Following \citet{2007MNRAS.381.1450N}, \citet{2014MNRAS.441..378L}, and \citet{2016MNRAS.457.4340K}, we adopt the stricter virial-ratio cut $2T/|U|<1.35$, which was introduced as a more conservative alternative to the commonly used $2T/|U|<1.5$ criterion given the uncertainties in estimating the virial parameter.

The relaxed fraction decreases steadily with redshift, reflecting the fact that halos are dynamically younger and more frequently disturbed at earlier times. A similar redshift dependence of halo relaxation indicators was also found by \citet{2016MNRAS.457.4340K}, who showed that the distributions of $x_{\rm off}$, $\lambda$, and $2T/|U|-1$ vary between $z=0$ and $z=3$. For the full sample, the relaxed fractions at $z=0$--5 are approximately 84\%, 80\%, 73\%, 63\%, 51\%, and 34\%, respectively. 
While our fiducial analysis focuses on relaxed halos, we also evaluate the complete, unselected halo sample (hereafter the ``all-halo'' sample) without applying dynamic relaxation cuts to evaluate the general applicability of our model across all evolutionary phases.

\subsection{Model formulation and parameterization}
\label{sec:variables}

Our central assumption is that the scatter in halo concentration is mainly driven by differences in halo formation history. The detailed procedures for computing $D(z)$, $\sigma(M)$, and $M_{\rm hm}$ are broadly based on \citetalias{2026ApJ..1006...61W}, with several key methodological refinements detailed below. 
In standard excursion set theory and halo models, the joint dependence on halo mass, redshift, and cosmology is often parameterized via the peak height $\nu(M,z) = \delta_{\rm sc}(z)/\sigma(M) = \delta_{\rm sc}(0)/[D(z)\sigma(M)]$, where $\delta_{\rm sc}(0)=1.686$ and $D(z)$ is the linear growth factor normalized to unity at $z=0$. The quantity $\sigma(M)$ is the linear density rms fluctuation on mass scale $M$ at $z=0$, computed from CAMB \citep{2000ApJ...538..473L}. Rather than forcing $D(z)$ and $\sigma(M)$ into a single composite variable $\nu$, in this work we decouple $D(z)$ and $\sigma(M)$ as independent regressors, allowing the empirical relation to capture background expansion and mass scale scalings with maximum flexibility.

Following \citetalias{2026ApJ..1006...61W}, the linear mass variance $\sigma(M)$ is computed using a real-space top-hat filter for CDM cosmologies. For WDM cosmologies, the sharp suppression of power below the free-streaming scale introduces artificial power leakage when evaluated with a real-space top-hat filter due to its $k^{-3}$ tail in Fourier space. To prevent this spurious small-scale contribution, we adopt a sharp-$k$ filter in Fourier space for WDM models (e.g., \citealp{2013MNRAS.433.1573S}; \citetalias{2026ApJ..1006...61W}).

To further encode the dependence on individual halo MAH, we utilize the EPS-scaled formation variable \citep{1993MNRAS.262..627L}
\begin{equation}
\omega_{\rm f} = \frac{\delta_{\rm sc}(z_{\rm f})-\delta_{\rm sc}(z)}{\sqrt{\sigma^2(fM)-\sigma^2(M)}},
\label{eq:omegaf}
\end{equation}
to serve as a proxy for halo formation epoch. Here $z_{\rm f}$ is defined as the redshift when the main progenitor first attains half (i.e. $f=1/2$) of the halo mass ($M_{\rm vir}$ or $M_{\rm 200c}$) at the redshift considered. In this work, $z_{\rm f}$ is measured from the main branch of the merger tree for each simulated halo. Compared with $z_{\rm f}$, $\omega_{\rm f}$ encodes implicit dependencies on mass, redshift, and cosmology within the excursion‑set framework, thus yielding a more uniform description of halo assembly across different simulations.

For WDM cosmologies, the suppression of small-scale power introduces an additional physical scale. We characterize this effect using the half-mode mass $M_{\rm hm}$, defined as the mass corresponding to the wavenumber $k_{\rm hm}$ at which the WDM transfer function is reduced by a factor of 2 relative to the corresponding CDM model with the same background cosmological parameters:
$M_{\rm hm}=\frac{4\pi}{3}\bar{\rho}\left(\frac{\pi}{k_{\rm hm}}\right)^3.$
In the empirical model below, this cutoff scale enters through
\begin{equation}
A_{\rm hm}=1+a\frac{M_{\rm hm}}{M_{\rm \Delta}}.
\label{eq:scut}
\end{equation}
Prior to performing the multivariate linear regression, the parameter $a$ is fixed to the values calibrated from the peak-concentration model (listed in Table~\ref{tab:fit}; see Appendix~\ref{app:peak_model} for details). For cosmologies without a WDM-like cutoff, $M_{\rm hm} = 0$ and hence $A_{\rm hm} = 1$.

To capture the joint dependence on cosmological expansion and power-spectrum fluctuation amplitude with maximum flexibility, we express the concentration of individual halos as
\begin{equation}
\begin{split}
\log c_{\rm \Delta} &= A + \alpha_{D} \log D(z) + \alpha_{\sigma} \log \sigma(M) \\
&\quad + \beta \log (1+\omega_{\rm f}) + \gamma \log A_{\rm hm}.
\end{split}
\label{eq:generalfit}
\end{equation}
Here, the scaling coefficients $\alpha_D$, $\alpha_\sigma$, $\beta$, and $\gamma$ capture the power-law dependencies on cosmic expansion, halo mass scale, assembly history, and the WDM cutoff suppression, respectively. We calibrate Equation~(\ref{eq:generalfit}) across the combined simulation suite via multiple linear regression, performed separately for relaxed halos and the complete all-halo sample.

We then test whether the concentration distribution can be predicted from the formation-time distribution. This is done in two ways. First, we combine Equation~(\ref{eq:generalfit}) with the EPS-predicted distribution of $\omega_{\rm f}$, which tests whether formation history scatter can explain concentration scatter statistically. Second, we replace the EPS prediction with $\omega_{\rm f}$ measured directly from simulated merger trees, which isolates the empirical concentration model from possible inaccuracies in EPS.

\section{Results}
\label{sec:results}

\subsection{Universal relation for individual halos}
\label{sec:universal_relation}

We first test whether Equation~(\ref{eq:generalfit}) can universally describe individual halo concentrations from $z = 5$ to $z = 0$. Because halos across our diverse simulation suites—spanning different cosmologies, volumes, and resolutions—follow a broadly consistent sequence, we combine all samples to perform a global multiple linear regression for both the relaxed and all-halo populations.

The best-fitting coefficients of Equation~(\ref{eq:generalfit}) for all combinations of dynamical states and halo definitions are summarized in Table~\ref{tab:fit}. Rather than writing out separate algebraic equations for each subset, Equation~(\ref{eq:generalfit}) serves as the master model relation, with specific parameters provided in Table~\ref{tab:fit}. The fitted relations share consistent functional dependencies on background expansion, density fluctuation amplitude, and assembly history. The coefficients $\alpha_D$ and $\alpha_\sigma$ are both positive, which is physically equivalent to a negative dependence on peak height ($\nu \propto [D\sigma]^{-1}$): halos with smaller mass (larger $\sigma$) and at lower redshift (larger $D$) have systematically higher concentrations. The positive coefficient of $\log(1+\omega_{\rm f})$ demonstrates that earlier-forming halos are more concentrated, directly confirming that concentration retains memory of the early collapse phase.

\begin{table*}[!htbp]
\centering
\caption{Best-fitting coefficients of the individual halo concentration model (Equation~\ref{eq:generalfit}) across two halo boundary definitions, for relaxed halos and the complete all-halo sample. The cutoff coefficient $a$ in $A_{\rm hm}$ is fixed to the peak-model values.}
\label{tab:fit}
\small
\begin{tabular*}{\textwidth}{@{\extracolsep{\fill}}l c c c c c c c c c c c c c c@{}}
\hline\hline
Sample &
\multicolumn{7}{c}{$\Delta_{\rm h}=\Delta_{\rm vir}$} &
\multicolumn{7}{c}{$\Delta_{\rm h}=200{\rm c}$} \\
\cline{2-8}\cline{9-15}
&
$A$ & $\alpha_D$ & $\alpha_\sigma$ & $\beta$ & $\gamma$ & $a$ & $\sigma_{\rm res}$ &
$A$ & $\alpha_D$ & $\alpha_\sigma$ & $\beta$ & $\gamma$ & $a$ & $\sigma_{\rm res}$ \\
\hline
Relaxed halos & 0.61 & 1.03 & 0.66 & 0.63 & -0.45 & 4.77 & 0.088 & 0.46 & 0.79 & 0.64 & 0.67 & -0.39 & 5.59 & 0.089 \\
All halos & 0.59 & 1.05 & 0.63 & 0.68 & -0.43 & 3.98 & 0.094 & 0.45 & 0.82 & 0.61 & 0.71 & -0.48 & 3.06 & 0.096 \\
\hline
\end{tabular*}
\end{table*}

\begin{figure*}
\centering
 \includegraphics[width=16.0cm,height=16.0cm]{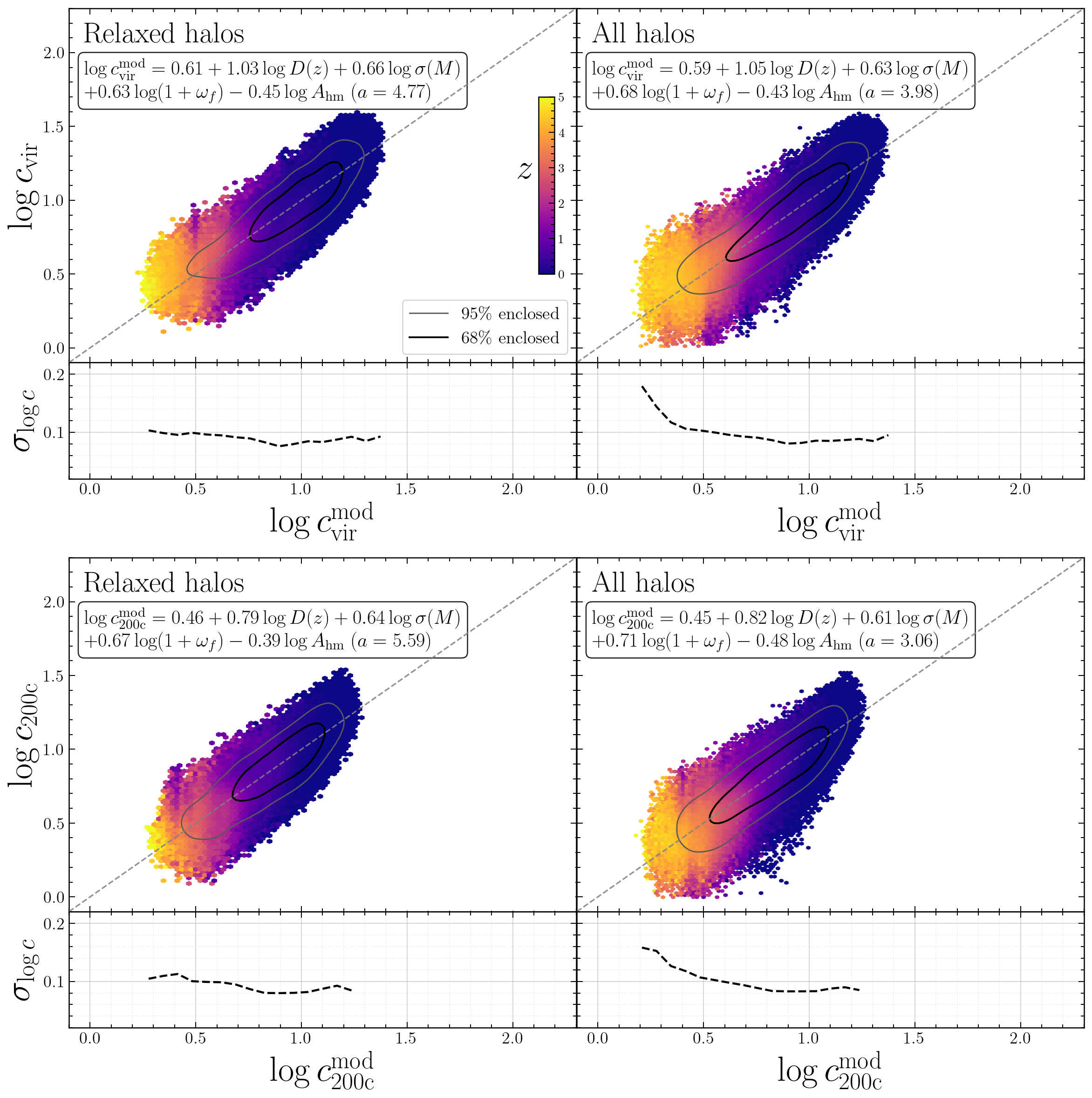}
\caption{Comparison between measured and predicted concentrations for the combined simulation dataset. The top row shows virial concentrations ($\Delta_{\rm h}=\Delta_{\rm vir}$), while the bottom row shows $200{\rm c}$ concentrations. The left column corresponds to relaxed halos, and the right column corresponds to the complete all-halo sample. Predictions, $\log c^{\rm mod}_{\Delta}$, are calculated using Equation~(\ref{eq:generalfit}) with the best-fitting coefficients from Table~\ref{tab:fit}. 
The distribution of individual halos is visualized via hexagonal binning color-coded by the mean redshift in each cell. 
Dashed lines indicate the one-to-one relation, while the contours enclose 95\% and 68\% of halos. Lower sub-panels display the binned dispersion, $\sigma_{\log c_{\Delta}}$, as a function of the model prediction.}
\label{Fig_all_sims}
\end{figure*}

Figure~\ref{Fig_all_sims} compares the measured concentrations with the regression predictions for both relaxed halos and the complete all-halo sample. The top row presents the results for $\Delta_{\rm h}=\Delta_{\rm vir}$, while the bottom row presents the results for $\Delta_{\rm h}=200{\rm c}$. Despite combining halos from diverse cosmologies, simulation volumes, resolutions, and redshifts, data points across all redshifts ($z=0$--5) cluster tightly around the one-to-one relation in all panels. This demonstrates that the physical parameters $D(z)$, $\sigma(M)$, and $\omega_{\rm f}$ successfully absorb the bulk of the explicit redshift, mass, and cosmological dependencies.

As shown in the left-column panels of Figure~\ref{Fig_all_sims},
the relations for relaxed halos remain remarkably tight across all redshifts, with a stable residual scatter of $\sigma_{\log c} \approx 0.08$--$0.10$ dex. In contrast, the complete all-halo sample (the panels in the right column) exhibits pronounced redshift evolution: its scatter is similarly narrow ($\sim 0.09$ dex) at $z=0$, but broadens significantly beyond $z=3$, reaching $\sigma_{\log c} \approx 0.18$ dex by $z=5$.
This elevated high-redshift dispersion is directly attributable to the dynamical state of the halo population discussed in Section~\ref{sec:relaxed_halo}: as the relaxed fraction drops precipitously from 84\% at $z=0$ to only 34\% at $z=5$, approximately two-thirds of high-$z$ halos are dynamically disturbed and undergoing violent mergers or rapid non-equilibrium accretion. These unrelaxed systems produce substantial transient scatter in halo internal structures and NFW profile fits, thereby driving up the high-redshift dispersion. Furthermore, as shown in Table~\ref{tab:fit}, the all-halo sample displays a slightly lower normalization $A$, reflecting the well-known tendency of unrelaxed halos to possess lower concentrations at fixed mass.

The comparison between the two halo definitions indicates that the $c_{200{\rm c}}$ relation is systematically shifted to lower concentrations than the $c_{\rm vir}$ relation, consistent with the smaller normalizations in Table~\ref{tab:fit}. This behavior is physically expected: $r_{200{\rm c}}$ probes an inner, denser region of the halo compared to $r_{\rm vir}$, so $M_{200{\rm c}} < M_{\rm vir}$ and $r_{200{\rm c}} < r_{\rm vir}$, yielding lower concentrations and a slightly modified sensitivity to background expansion. 

Ultimately, Equation~(\ref{eq:generalfit}) provides a simple empirical description of individual halo concentrations. An individual halo's concentration is primarily governed by three fundamental factors: cosmic background expansion and initial power spectrum amplitude, assembly history, and, where relevant, the WDM cutoff scale in the initial power spectrum.
From this equation, one can explain the dependence of individual‑halo concentration on redshift, halo mass, the initial power spectrum, cosmology, and halo MAH. The dependencies on redshift and initial power‑spectrum are similar to those presented in \citetalias{2026ApJ..1006...61W}: halos at higher redshift exhibit smaller concentrations because $D(z)$ decreases with increasing $z$, while halos in cosmologies with a larger initial power‑spectrum normalization $\sigma_{8}$ (and hence larger $\sigma(M)$) yield larger concentrations. Meanwhile, more‑massive halos have smaller concentrations, given that $\sigma(M)$ decreases with $M$. 
However, unlike the formulae for peak‑halo concentration presented in \citetalias{2026ApJ..1006...61W}, the equation above quantitatively captures the relation with individual halo MAH. As expected, a halo formed at an earlier epoch (with larger $z_{\rm f}$ and hence larger $\omega_{\rm f}$) possesses a larger concentration.
The above dependencies can be understood theoretically in such a way that a higher $\sigma_8$ and an earlier individual formation epoch $z_{\rm f}$ both imply that the halo’s inner core collapsed at an earlier cosmic time when the background universe was denser, imprinting a more compact scale radius $r_s$ and hence a higher concentration. Conversely, toward lower redshift $z$, the drop in cosmic background density coupled with late-time envelope growth expands the outer radius $r_{\Delta}$ relative to the stabilized inner core, systematically increasing $c_\Delta = r_{\Delta} / r_{\rm s}$. Dependencies on other cosmological parameters are more complex, as they are embedded within the relations for $D(z)$ and $\sigma(M)$.

\begin{figure*}
\centering
 \includegraphics[width=\textwidth]{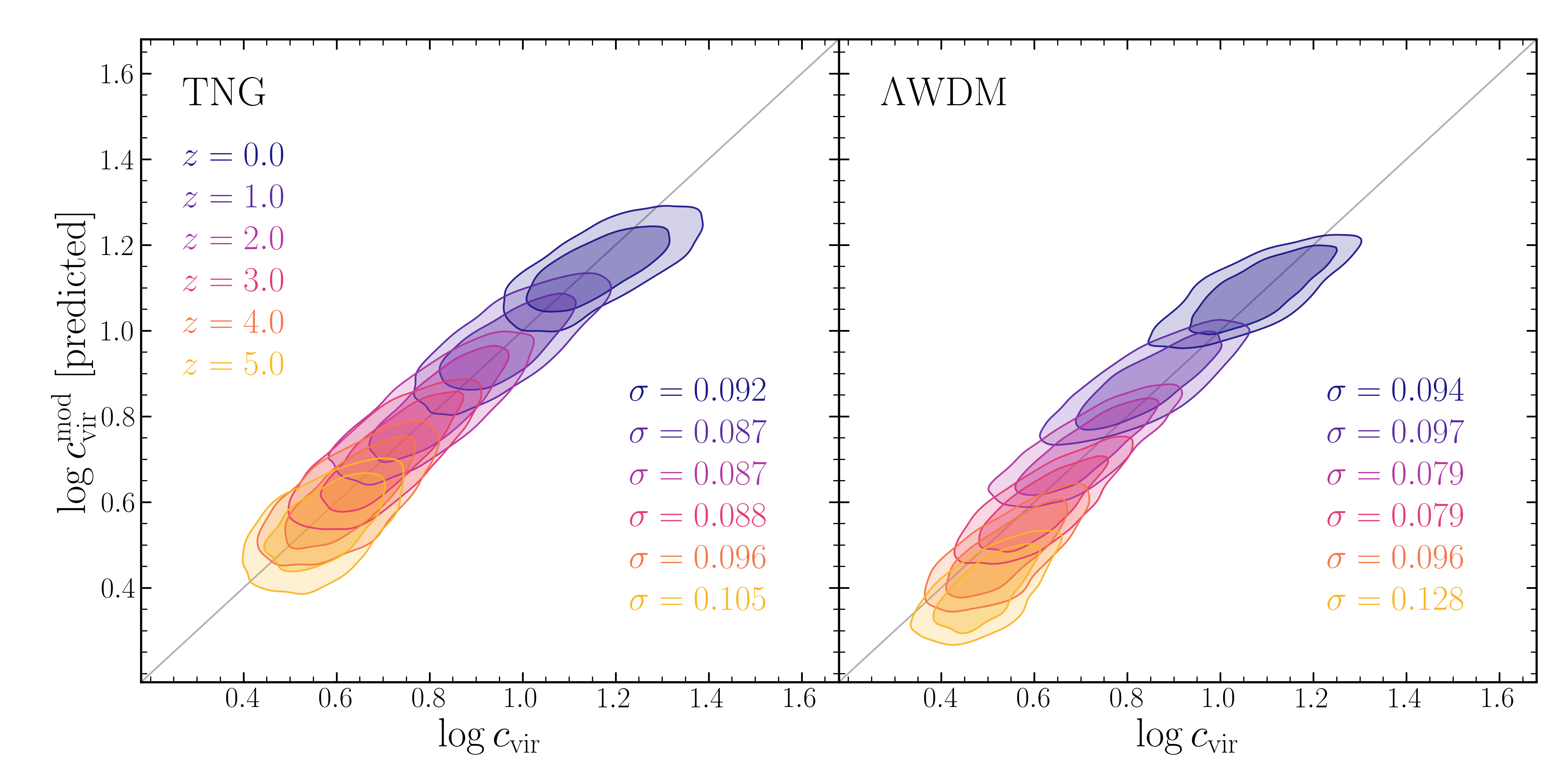}
\caption{
Comparison of the predicted and measured concentrations for relaxed halos in the TNG ($\Lambda$CDM, left) and $\Lambda$WDM (right) simulations.}
Different colors represent different redshifts. For each redshift, the outer and inner contours enclose 75\% and 50\% of the halos, respectively. The grey diagonal line marks the one-to-one relation. Across both cosmologies, the scatter around equality remains remarkably tight at $\sigma \approx 0.09$ dex.
\label{fig:c_comparison}
\end{figure*}

While Figure~\ref{Fig_all_sims} demonstrates the overall tightness of the global regression across the pooled dataset, combining all simulations and epochs into a single plot can obscure potential systematic biases within specific cosmological models or individual redshifts. To verify that Equation~(\ref{eq:generalfit}) holds unbiasedly across distinct regimes without recalibration, we apply the globally calibrated relation (with the fixed coefficients in Table~\ref{tab:fit}) to individual relaxed halos in the TNG ($\Lambda$CDM) and $\Lambda$WDM simulations at separate redshifts. Figure~\ref{fig:c_comparison} presents the predicted versus the measured concentrations for these two simulation suites.

For both cosmologies, halos at different redshifts follow a tight sequence close to the one-to-one line almost without systematic tilts. The colored contours enclose 75\% and 50\% of the halos at each redshift, showing that the bulk of the population is accurately reproduced by the model. 
The typical scatter about the one‑to‑one relation remains remarkably stable at $\sigma \lesssim 0.1\,\mathrm{dex}$ for both $\Lambda$CDM and $\Lambda$WDM across nearly all tested redshifts, apart from the cases at $z=5$.

The tight agreement in the $\Lambda$WDM case is especially noteworthy. Although the suppression of small-scale power substantially alters halo accretion histories and lowers the concentrations of low-mass halos, our cutoff parameter $A_{\rm hm}$, with $a$ fixed to the calibrated peak-model value, successfully absorbs this suppression. This indicates that individual halo concentration variations across cosmologies can be reliably modeled by combining linear variance, scaled formation epoch, and cutoff scale. With the individual relation verified, we now turn to generating the halo concentration distributions via Monte Carlo sampling.

\subsection{Predicting concentration distributions from formation histories}
\label{sec:predict_c}

The individual halo relation in Equation~(\ref{eq:generalfit}) can be utilized not only to predict the concentration of a given halo, but also to reconstruct the concentration distribution at fixed mass and redshift. If the scatter in concentration is predominantly driven by halo assembly history, then a distribution of formation variables, $P(\omega_{\rm f})$, should map directly into the concentration distribution via Monte Carlo realizations.

\begin{figure*}[!htbp]
\centering
\includegraphics[width=\textwidth]{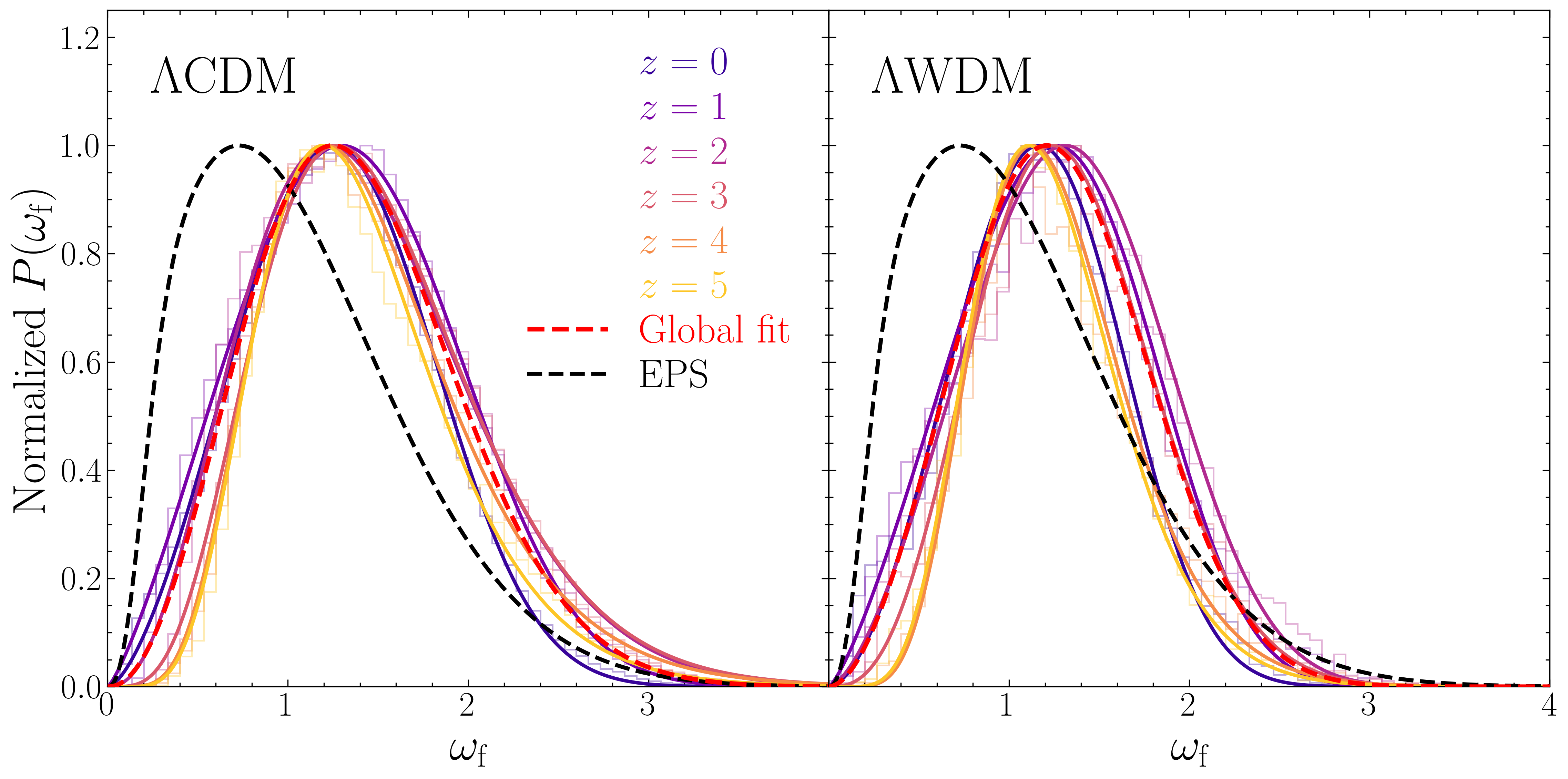}
\caption{Normalized probability density functions (PDFs) of the scaled formation variable $\omega_{\rm f}$ from $z=0$ to $z=5$. The left and right panels show the numerical simulation measurements (colored histograms and curves) for $\Lambda$CDM and $\Lambda$WDM, respectively. These are compared with the theoretical expectation from standard EPS theory at $z=0$ (dashed black curves). The global fits across all redshifts using the generalized gamma distribution (Equation~\ref{eq:gengamma_fit}) are displayed as red dashed curves. Notably, the simulated distributions peak at significantly higher values than standard EPS (highlighting a delayed-formation bias in EPS theory), and the WDM distribution is markedly narrower due to the suppression of small-scale perturbations.}
\label{fig:omega_f_compare}
\end{figure*}

To inspect the input distribution, Figure~\ref{fig:omega_f_compare} compares the probability density functions (PDFs) of $\omega_{\rm f}$ measured directly from cosmological simulations across redshifts ($z=0$--5) with the theoretical expectation from standard EPS theory at $z=0$ (dashed black curves). Two key physical insights emerge from this comparison. First, standard EPS systematically predicts later halo formation times, peaking at $\omega_{\rm f, peak}^{\rm EPS} \approx 0.75$ (as derived in the Appendix of \citetalias{2026ApJ..1006...61W}), significantly lower than the actual peaks measured in the simulations (see Figure~\ref{fig:omega_f_compare}).
Because $\omega_{\rm f}$ is defined such that larger values correspond to earlier collapse epochs (Equation~\ref{eq:omegaf}), this systematic deficit directly explains why EPS theory underestimates halo formation times relative to $N$-body simulations \citep[e.g.,][]{2003MNRAS.344.1327L}. Second, the $\omega_{\rm f}$ distribution in WDM is markedly narrower than in $\Lambda$CDM. This occurs because the primordial power spectrum in WDM is severely suppressed below the half-mode scale $M_{\rm hm}$ due to dark matter thermal free-streaming. In the absence of small-scale perturbations, hierarchical, stochastic mergers of tiny sub-components are extinguished; instead, halos near the truncation scale form delayed, rapidly, and monolithically \citep[e.g.,][]{2001ApJ...556...93B,2012MNRAS.424..684S,2016MNRAS.460.1214L}. Because their non-linear collapse is tightly coupled to the epoch when the cutoff scale enters the non-linear regime, halo assembly histories in WDM become remarkably homogeneous, compressing the distribution into a substantially narrower width.

A closer inspection of Figure~\ref{fig:omega_f_compare} indicates minor redshift variations: with increasing redshift, the simulated distributions shift slightly toward lower $\omega_{\rm f}$ and become marginally narrower,
reflecting younger dynamical states and shorter available accretion timescales at early epochs. Nonetheless, the distributions maintain broad cosmological self-similarity across cosmic time. By definition (Equation~\ref{eq:omegaf}), the scaled formation variable is non-negative ($\omega_{\rm f} \ge 0$), naturally vanishing at $\omega_{\rm f} = 0$. We therefore parameterize the overall distribution of $\omega_{\rm f}$ across all redshifts using a generalized gamma distribution \citep{stacy1962generalization, 2007MNRAS.376..215B}:
\begin{equation}
P(\omega_{\rm f}) = C_0 \left(\frac{\omega_{\rm f}}{\omega_0}\right)^k \exp\left[ -\frac{k}{p} \left(\frac{\omega_{\rm f}}{\omega_0}\right)^p \right], \quad (\omega_{\rm f} \ge 0),
\label{eq:gengamma_fit}
\end{equation}
where $\omega_0 \equiv \omega_{\rm f, peak}$ directly corresponds to the peak of the distribution, $k$ governs the low-$\omega_{\rm f}$ power-law rise, $p$ controls the exponential cutoff, and the normalization constant is given by
\begin{equation}
C_0 = \frac{p}{\omega_0} \frac{(k/p)^{(k+1)/p}}{\Gamma\left(\frac{k+1}{p}\right)},
\end{equation}
where $\Gamma(x) = \int_0^\infty t^{x-1} e^{-t} \, \mathrm{d}t$ is the standard gamma function. Fitting Equation~(\ref{eq:gengamma_fit}) across all redshifts yields $(\omega_0, k, p) = (1.25, 2.14, 2.04)$ for $\Lambda$CDM and $(1.21, 1.82, 2.70)$ for $\Lambda$WDM (shown as red dashed curves in Figure~\ref{fig:omega_f_compare}), corresponding to simulated peak values of $\omega_{\rm f, peak} \approx 1.25$ and $\approx 1.21$, respectively. For simplicity, we adopt this global fit across cosmic epochs, as the minor redshift dependencies have negligible impact on the reconstructed concentration statistics.

Equation~(\ref{eq:gengamma_fit}) shares deep physical asymptotes with the theoretical EPS distribution, $p(\omega_{\rm f}) = 2\omega_{\rm f} \, \mathrm{erfc}(\omega_{\rm f}/\sqrt{2})$ (\citealp{1993MNRAS.262..627L}; \citetalias{2026ApJ..1006...61W}). In the rare-peak, early-formation limit ($\omega_{\rm f} \gg 1$), the complementary error function asymptotically expands as $\mathrm{erfc}(x) \sim e^{-x^2}/(\sqrt{\pi} x)$, yielding an exponential tail $p(\omega_{\rm f}) \propto \exp(-\omega_{\rm f}^2/2)$ governed by the Gaussian statistics of the linear density field. The exponential tail in Equation~(\ref{eq:gengamma_fit}) faithfully preserves this physical asymptote, with the fitted exponent $p \approx 2.04$ in $\Lambda$CDM closely recovering the theoretical Gaussian decay $p=2$. In the low-$\omega_{\rm f}$ regime ($\omega_{\rm f} \to 0$), while standard EPS features a linear rise ($p(\omega_{\rm f}) \propto \omega_{\rm f}$) driving a steep macroscopic slope toward its early theoretical peak ($\omega_{\rm f, peak}^{\rm EPS} \approx 0.75$), the simulation distributions exhibit a quadratic suppression ($k \approx 1.82\text{--}2.14$) and a broader rise toward a delayed peak. In addition, the steeper exponential tail in WDM ($p \approx 2.70$) cleanly captures the enhanced cutoff suppression and the synchronized, rapid collapse of structures near the free-streaming scale.

We then perform Monte Carlo realizations to reproduce the concentration distribution from the formation history PDF $P(\omega_{\rm f})$. For each halo mass and redshift (which fix the proxies $D(z)$ and $\sigma(M)$, as well as $A_{\rm hm}$), random values of $\omega_{\rm f}$ are drawn from the PDF $P(\omega_{\rm f})$ and substituted into the empirical regression relation (Equation~\ref{eq:generalfit}) to synthesize individual halo concentrations and predict the $c(M)$ distribution.

\begin{figure*}
\centering
\includegraphics[width=\textwidth]{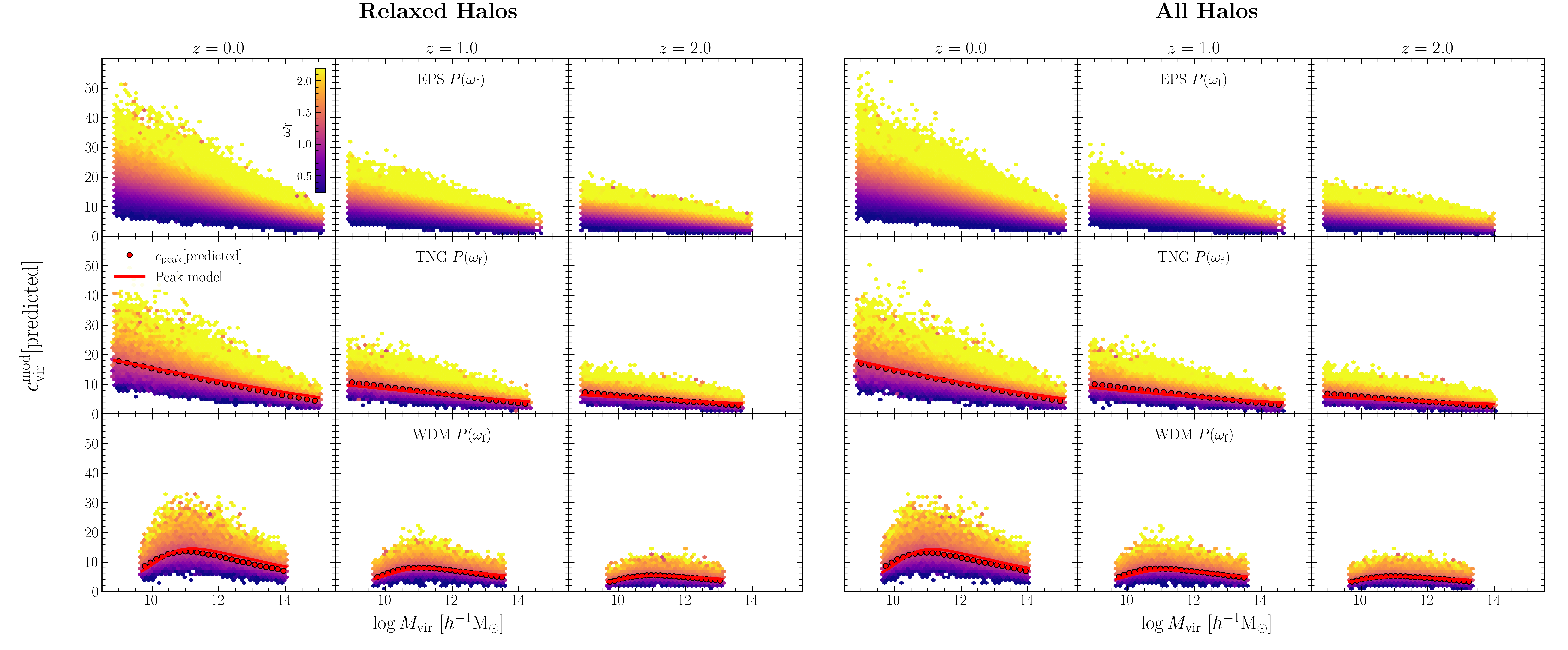}
\caption{Monte Carlo predictions of concentration--mass distributions at $z = 0$--2, generated by substituting random $\omega_{\rm f}$ draws into Equation~(\ref{eq:generalfit}). Left and right columns show relaxed and all halos, respectively. Rows indicate the $\omega_{\rm f}$ input source: theoretical EPS for $\Lambda$CDM (top), simulation PDF for $\Lambda$CDM (middle), and simulation PDF for $\Lambda$WDM (bottom). Solid red curves display the peak concentration model (from \citetalias{2026ApJ..1006...61W} for all halos; Appendix~\ref{app:peak_model} for relaxed halos), and red circles mark the extracted binned lognormal peak values. The color scale denotes $\omega_{\rm f}$.}
\label{fig:cdis_prediction}
\end{figure*}

The upper panels of Figure~\ref{fig:cdis_prediction} show the resulting Monte Carlo predictions when drawing random $\omega_{\rm f}$ from the theoretical EPS distribution for TNG at $z=0$--2. While the EPS-based model captures the overall qualitative trends, it systematically underpredicts halo concentrations, especially at the low-mass end. This offset arises directly from the underestimation of $\omega_{\rm f}$ by EPS theory.

We then repeat this Monte Carlo reproduction using random $\omega_{\rm f}$ drawn directly from the simulation-measured PDF (Equation~\ref{eq:gengamma_fit}). The middle and lower panels of Figure~\ref{fig:cdis_prediction} show that these simulation-based Monte Carlo predictions eliminate this offset and track the peak concentration--mass relation of \citetalias{2026ApJ..1006...61W} (red lines, corresponding to \citetalias{2026ApJ..1006...61W} for all halos and our Appendix~\ref{app:peak_model} calibration for relaxed halos) and the extracted binned lognormal peak values (red circles) remarkably well across all masses and redshifts in both $\Lambda$CDM and $\Lambda$WDM. In WDM cosmologies (bottom row), incorporating simulation-drawn $\omega_{\rm f}$ along with $A_{\rm hm}$ accurately reproduces the flattening and turnover in halo concentrations below $M_{\rm hm}$.

\subsection{Comparison with other models}
\label{sec:comparison_model}

To assess the performance and applicability of our empirical formulation, we compare our individual halo predictions with previous concentration models, most notably the physically motivated model developed by \citet{2016MNRAS.460.1214L}. The Ludlow et al.\ model relates the internal characteristic density of a dark matter halo to the background cosmic critical density at the collapse redshift $z_{-2}$, defined as the epoch when the cumulative collapsed mass history (CMH) of the main progenitor first assembled the core mass $M_{-2}$. In their work, halo density profiles were modeled using an Einasto \citep{1965TrAlm...5...87E} profile with a fixed shape parameter of $\alpha = 0.18$, and concentrations were defined relative to the cosmic critical density as $c_{200{\rm c}} = r_{200{\rm c}}/r_{-2}$.

\begin{figure}
\centering
\includegraphics[width=\columnwidth]{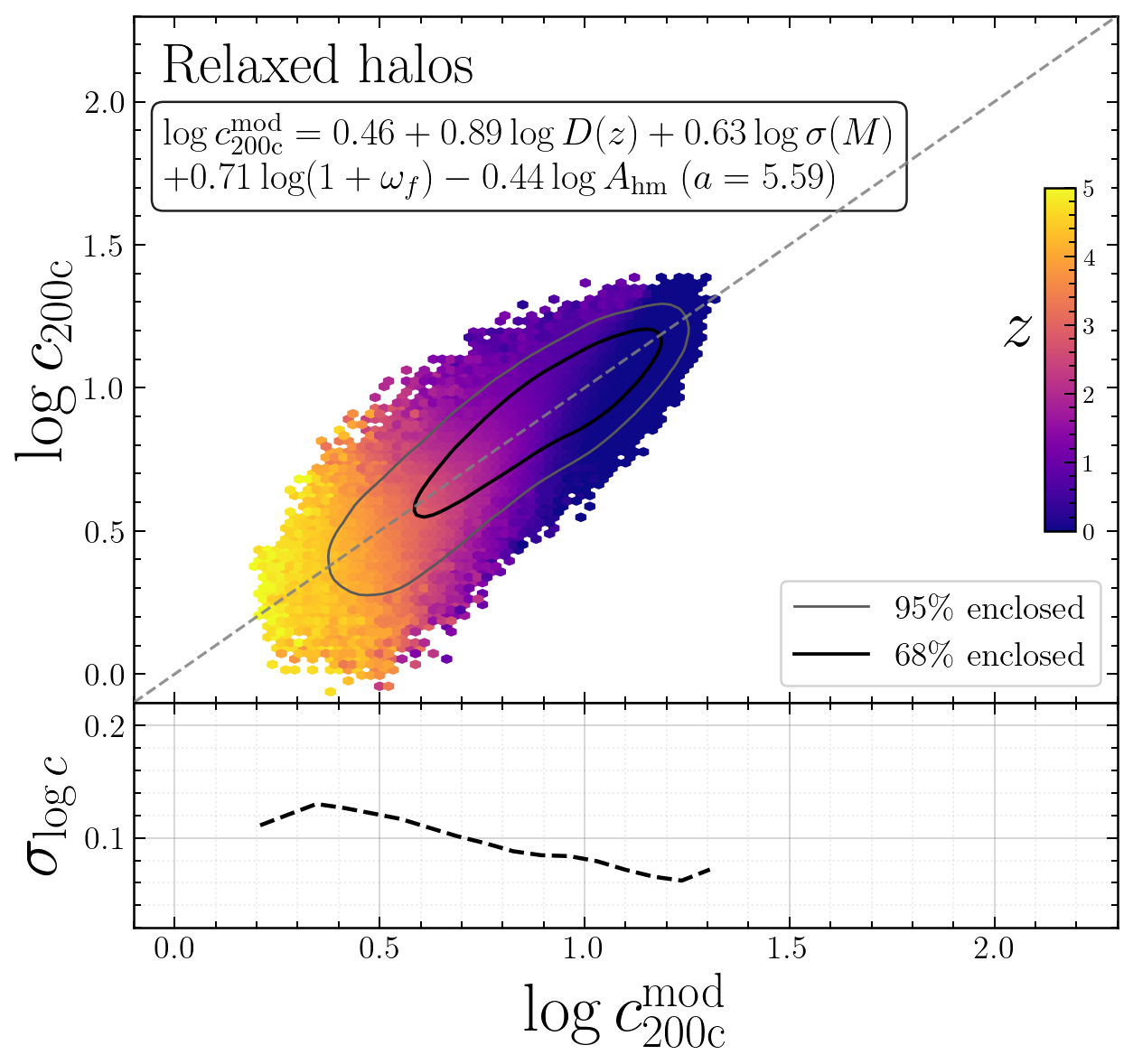}
\caption{Global fit of our empirical concentration model for relaxed halos under the $\Delta_{\rm h}=200{\rm c}$ definition, calibrated using concentrations derived from an Einasto profile with $\alpha = 0.18$. 
The distribution combines individual halos from the TNG and non-$\Lambda$CDM simulations from $z = 5$ to $z = 0$, visualized via hexagonal binning color-coded by the mean redshift in each cell.}
The dashed line marks the one-to-one diagonal, and contours enclose 95\% and 68\% of halos. The lower sub-panel shows the binned residual scatter, achieving an average dispersion of $\sigma_{\log c} \approx 0.098$ dex.
\label{fig:einasto_fit}
\end{figure}

To ensure a rigorous and direct comparison, we recalibrate our empirical multi-linear relation (Equation~\ref{eq:generalfit}) using the exact same profile definition: concentrations derived by fitting an Einasto profile with fixed $\alpha = 0.18$ within $r_{200{\rm c}}$. Furthermore, because \citet{2016MNRAS.460.1214L} evaluated linear density fluctuations using a real-space top-hat filter for all cosmologies (including WDM), in this subsection we also evaluate $\sigma(M)$ using the real-space top-hat filter to maintain an identical baseline. 
Because full dark matter particle snapshots across all redshifts are not publicly available for the MultiDark-Planck and Uchuu suites, we restrict this Einasto profile recalibration to relaxed halos from the TNG and our non-$\Lambda$CDM simulations.
The resulting Einasto-based empirical relation is given by
\begin{equation}
\begin{split}
\log c_{200{\rm c}}^{\rm Einasto} &\approx 0.46 + 0.89 \log D(z) + 0.63 \log \sigma(M) \\ 
&\quad + 0.71 \log (1+\omega_{\rm f}) - 0.44 \log A_{\rm hm}
\end{split}
\label{eq:einasto_fit}
\end{equation}
with the WDM cutoff parameter fixed to $a = 5.59$ as determined from the relaxed peak model under the $200{\rm c}$ boundary definition. Figure~\ref{fig:einasto_fit} shows the regression performance of Equation~(\ref{eq:einasto_fit}) across the TNG and non-$\Lambda$CDM cosmological simulation datasets. The predicted concentrations tightly follow the measured Einasto values across the entire range $z=0$--5, exhibiting an exceptionally small residual scatter that increases from $\sigma_{\log c} \approx 0.06$ dex at $z=0$ to $\approx 0.13$ dex at $z=5$. This confirms that our four-variable formulation accurately captures halo concentrations regardless of whether an NFW or an Einasto density profile is adopted.

\begin{figure*}
\centering
 \includegraphics[width=\textwidth]{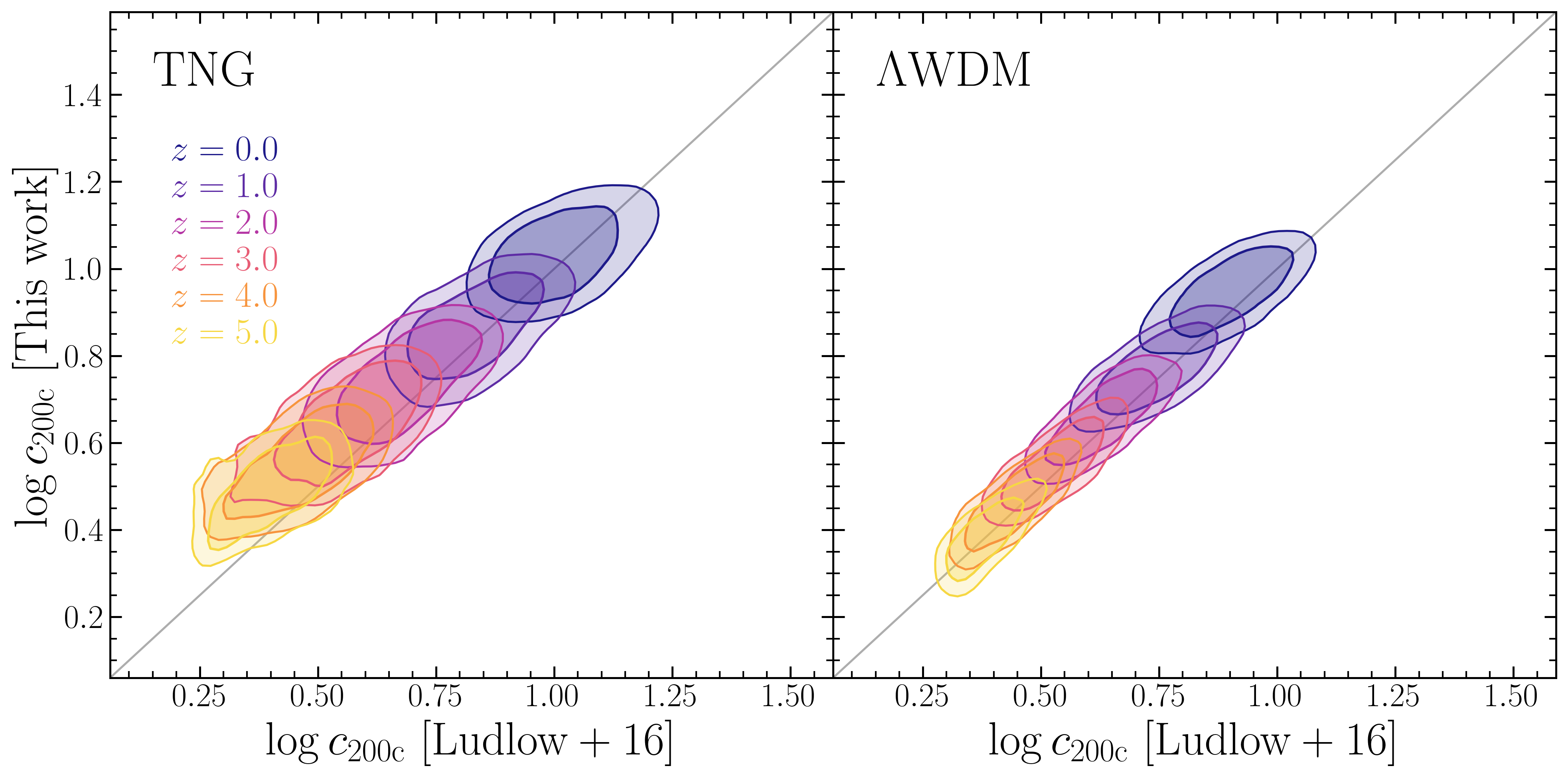}
\caption{
Comparison of the $200{\rm c}$ concentrations predicted by our Einasto-calibrated empirical model (Equation~\ref{eq:einasto_fit}) and those predicted by the model of \citet{2016MNRAS.460.1214L} for relaxed halos in the TNG ($\Lambda$CDM, left) and $\Lambda$WDM (right) simulations. }
Different colors represent redshifts from $z=0$ to $z=5$. Contours enclose 75\% and 50\% of the halo population, while the grey diagonal lines denote equality. In both cosmological frameworks, the two independent models demonstrate remarkable broad consistency across the entire dynamic range from $z=0$ to $z=5$.
\label{fig:model_comparison}
\end{figure*}

In \citet{2016MNRAS.460.1214L} (their Figure~10), the authors evaluated their model against numerical simulations (the COCO-COLD and COCO-WARM suites) only across three low-to-intermediate redshifts: \(z=0\)–2. Over this redshift baseline ($z \le 2$), their model was shown to reproduce individual halo concentrations with an rms scatter of $\sigma \lesssim 0.09$ dex. However, how their model behaves at higher redshifts ($z > 2$) was not presented in their original paper. Here, we extend the comparison between the two models across all redshifts ($z = 0$--$5$) for both $\Lambda$CDM (TNG) and $\Lambda$WDM simulations, as presented in Figure~\ref{fig:model_comparison}.

As shown in Figure~\ref{fig:model_comparison}, the two independently developed models show broad and remarkable consistency across both cosmologies. Over the redshift interval $z = 0$--2 where the Ludlow et al.\ model was originally calibrated, the predictions of both models are in excellent agreement, closely aligning along the one-to-one diagonal. At higher redshifts ($z = 3$--5), both models continue to track the same overall sequence over several orders of magnitude in halo mass. In $\Lambda$WDM (right panel), the two models remain remarkably consistent along equality all the way to $z = 5$. In $\Lambda$CDM (left panel), the contours exhibit a modest shift at $z \ge 3$, where our model predicts slightly higher concentrations by $\sim 0.08$--$0.10$ dex. This difference remains comparable to the typical halo-to-halo intrinsic scatter ($\sim 0.15$--$0.20$ dex) at these early epochs, confirming that both frameworks capture the primary physical drivers of halo internal structure.

While both approaches demonstrate that halo assembly history predominantly governs halo concentration, they represent complementary modeling philosophies. The Ludlow et al.\ model is grounded in a physical linkage between inner density and the critical density at the CMH crossing epoch, which requires tracing the complete hierarchy of progenitor merger branches and numerically solving a transcendental equation for every individual halo. In contrast, our empirical relation provides an explicit, closed-form algebraic formula that requires only four readily accessible parameters ($D(z)$, $\sigma(M)$, $\omega_{\rm f}$, and $A_{\rm hm}$). This compact mathematical structure makes our model computationally lightweight and straightforward to integrate into large-scale semi-analytic galaxy formation models and cosmological mock catalogs.

We also compare our results with the machine-learning approach of \citet{2025Univ...11...37Z}, who trained neural networks to predict halo concentrations from complete MAHs over $z=0$--2. While their neural network achieves an RMSE of 0.0845 at $z=0$, the scatter increases to 0.1287 at $z=2$, and predictions beyond redshift 2 were not provided. Our simple analytical relation achieves a competitive, nearly uniform scatter across $z=0$--5 with clear physical interpretability, providing a transparent and robust framework for individual halo concentration predictions across diverse cosmological scenarios.

\section{Discussion and Conclusions}
\label{sec:discussion}

In this work, we have established a concise, empirical formulation for predicting the internal concentrations of individual dark matter halos across both cold and warm dark matter cosmologies spanning from redshift $z=5$ down to $z=0$. By explicitly decoupling cosmological expansion via the linear growth factor $D(z)$, mass scale through the linear fluctuation amplitude $\sigma(M)$, individual mass accretion histories through the EPS-scaled formation variable $\omega_{\rm f}$, and the small-scale damping scale via $A_{\rm hm}$, our model accurately captures the halo-to-halo concentration scatter and its cosmological evolution.

Calibrated against an extensive suite of cosmological $N$-body simulations, our model achieves a typical residual dispersion of $\approx 0.08$--$0.10$ dex for relaxed halos across all examined redshifts. For the complete, unselected halo population, the scatter remains similarly narrow ($\approx 0.09$ dex) at $z=0$ but exhibits pronounced redshift evolution, increasing to $\approx 0.18$ dex at $z=5$ due to the rising fraction of dynamically disturbed systems at early epochs. Both virial and $200{\rm c}$ boundary definitions are well described by this framework. When applied to reconstruct concentration distributions, adopting $\omega_{\rm f}$ measured directly from simulation merger trees faithfully reproduces both the empirical distributions and the peak-concentration relations from our previous work. In contrast, standard EPS theory systematically underpredicts concentrations due to an inherent delayed-formation bias.

In summary, using $D(z)$, $\sigma(M)$, and $\omega_{\rm f}$ as proxies for halo-identification redshift, halo mass, and halo assembly epoch, respectively, our model can accurately predict the concentration parameter for individual halos given their MAH. The dependencies on redshift, halo mass, cosmology, the initial power‑spectrum and MAH are embedded within these three proxies.
Compared with the model of \citet{2016MNRAS.460.1214L}, which requires tracking the full progenitor‑merger trees of the target halo, our formulation provides comparable predictive fidelity via an explicit closed‑form algebraic relation that is also computationally convenient and has been extended to redshift 5. This makes our framework particularly well‑suited for rapid implementation within large‑scale cosmological mock pipelines and semi‑analytic galaxy‑formation models. By coupling our closed‑form relation with merger‑tree formation epochs or Monte‑Carlo draws from our calibrated \(P(\omega_{\rm f})\) distribution, galaxy‑formation models and survey mock catalogs can self‑consistently propagate halo assembly histories into internal density profiles, galaxy structural scaling relations, and predictions for small‑scale clustering or lensing.
Conversely, given observational measurements of halo mass and concentration parameters, for example, from lensing or dynamical analysis, our model enables us to infer the merger history of the target halo.
A user-friendly code package implementing the concentration models of \citetalias{2026ApJ..1006...61W} and this work is publicly available on GitHub.\footnote{\url{https://github.com/wangdz630/concentration_universal_model/}}

Although our empirical formulation accounts for the primary driver of concentration diversity and stabilizes the typical scatter to $\approx 0.08$--$0.10$ dex for relaxed halos, a non-negligible fraction of the variance remains unexplained, since mass‑accretion histories alone reproduce only 40\%–50\% of the total scatter within the CDM framework (c.f., \citealp{2016MNRAS.460.1214L}). Other plausible contributors to halo‑structure variance include environmental modulations such as large-scale tidal fields and anisotropic shear. Our framework offers a natural pathway to investigate how large-scale environment modulates halo structure. Because halos in different cosmic web environments (e.g., nodes, sheets, filaments and voids) experience systematically different assembly histories—closely linked to halo assembly bias (e.g., \citealp{2008MNRAS.387..536G})—our model can quantify how environment-dependent shifts in the conditional $\omega_{\rm f}$ distribution drive concentration variations. 
Disentangling how the cosmic environment shapes halo internal profiles both through and beyond their mass accretion histories represents a promising direction for future work.

\begin{acknowledgments}
The authors thank the TNG, MultiDark-Planck, Uchuu projects and 
Dr. Peter Behroozi for publicly releasing the Rockstar and Consistent trees codes.
This research has been supported by the NSFC grants (Nos. 12573007, 12073089). WSZ is supported by National SKA Program of China (grant Nos. 2025SKA0150104) and NSFC grant 12595314.
Parts of the simulations and all of the analysis in this work were carried out at the Kunlun \& Loong HPC facility of the School of Physics and Astronomy, Sun Yat-sen University.
\end{acknowledgments}

\appendix
\section{Calibration of the Peak Concentration Model for Relaxed Halos}
\label{app:peak_model}

In Section~\ref{sec:predict_c} (and Figure~\ref{fig:cdis_prediction}), we test whether our individual halo concentration relation (Equation~\ref{eq:generalfit}) can statistically reconstruct the global concentration distribution when combined with the assembly history distribution $P(\omega_{\rm f})$. To rigorously evaluate whether the predicted concentration sequences match the typical halo population across mass, redshift, and cosmology, an independent population-level benchmark is required. For this purpose, we benchmark our reconstructed distributions against the universal peak-concentration framework established in \citetalias{2026ApJ..1006...61W}.

Furthermore, while \citetalias{2026ApJ..1006...61W} calibrated the peak-concentration relation exclusively for the complete, unselected halo population (the all-halo sample), this work places primary emphasis on relaxed halos. Therefore, to provide the exact reference benchmark for relaxed halos in Figure~\ref{fig:cdis_prediction} and to physically ground the WDM cutoff parameter $a$ in Equation~(\ref{eq:scut}), here we summarize the peak model formulation and report its newly calibrated parameters for relaxed halos.

\subsection{Model Formulation}

Following \citetalias{2026ApJ..1006...61W}, the peak concentration of halos is characterized by their collapse history through the formation peak-height framework. The characteristic linear growth factor evaluated at the peak formation epoch, $D(z_{\rm f,peak})$, is decomposed into a baseline CDM component and a WDM suppression factor:
\begin{equation}
D(z_{\rm f,peak}) = D(z_{\rm f,peak})_{\rm CDM} \, \mathcal{E}(M_{\rm \Delta},M_{\mathrm{hm}}),
\label{eq:app_D_zf_peak}
\end{equation}
where the baseline CDM growth factor follows
\begin{equation}
D(z_{\rm f,peak})_{\rm CDM} = D(z) \left[ a_1 \left(\frac{M_{\rm \Delta}}{10^{12}h^{-1}\mathrm{M_\odot}}\right)^{b_1} + a_2 D(z)^{b_2} \right],
\label{eq:app_Dzf_cdm}
\end{equation}
and the WDM modification factor is parameterized as
\begin{equation}
\mathcal{E}(M_{\rm \Delta},M_{\mathrm{hm}}) = 1.0 + \eta \left( \frac{M_{\rm hm}}{M_{\rm \Delta}} \right)^\mu.
\label{eq:app_EM_peak}
\end{equation}
Here $M_{\rm hm}$ is the half-mode mass of the WDM model (with $M_{\rm hm} = 0$ and $\mathcal{E} = 1$ for cosmologies without a WDM-like cutoff).

The peak concentration $c_{\rm peak}$ is then universally described as a power law of the effective formation peak height $\nu_{\rm eff}$:
\begin{equation}
c_{\rm peak} = A_{\rm p} \nu_{\rm eff}^{-\alpha_{\rm p}} + c_{\rm 0},
\label{eq:app_c_peak_model}
\end{equation}
where $\nu_{\rm eff}$ incorporates the small-scale cutoff suppression via
\begin{equation}
\nu_{\rm eff} = \nu_{\rm peak} A_{\rm hm}, \qquad \nu_{\rm peak} \equiv \frac{\delta_{\rm sc}(0)}{D(z) D(z_{\rm f,peak}) \sigma(M)}.
\label{eq:app_nu_eff}
\end{equation}

\subsection{Calibrated Parameter Sets}

For the complete all-halo sample, the parameters were established in \citetalias{2026ApJ..1006...61W} using the sharp-$k$ filter for WDM. For relaxed halos, we apply the exact same calibration procedure to the relaxed halo catalogs across all our cosmological simulations from $z=5$ to $z=0$. 

The resulting best-fitting parameter sets for both relaxed halos and the complete all-halo population, under both the virial ($\Delta_{\rm vir}$) and 200 times critical density ($\Delta_{200c}$) boundary definitions, are summarized in Table~\ref{tab:app_peak_params}.

\begin{table*}[!htbp]
\centering
\caption{Best-fitting parameters of the formation peak-height concentration model across two halo definitions ($\Delta_{\rm vir}$ and $\Delta_{200c}$). Parameters for the all-halo sample are taken from \citetalias{2026ApJ..1006...61W}, while those for relaxed halos are newly calibrated in this work.}
\label{tab:app_peak_params}
\small
\begin{tabular*}{\textwidth}{@{\extracolsep{\fill}}l c cccc cc ccc c@{}}
\hline\hline
Sample & Boundary & \multicolumn{4}{c}{Baseline $D(z_{\rm f,peak})_{\rm CDM}$} & \multicolumn{2}{c}{WDM Factor $\mathcal{E}$} & \multicolumn{3}{c}{Peak Relation $c_{\rm peak}(\nu_{\rm eff})$} & Cutoff \\
\cline{3-6}\cline{7-8}\cline{9-11}\cline{12-12}
& $\Delta_{\rm h}$ & $a_1$ & $b_1$ & $a_2$ & $b_2$ & $\eta$ & $\mu$ & $A_{\rm p}$ & $\alpha_{\rm p}$ & $c_0$ & $a$ \\
\hline
Relaxed halos & $\Delta_{\rm vir}$ & 0.21 & 0.10 & 0.33 & $-0.48$ & 0.14 & 1.07 & 13.20 & 0.93 & 2.10 & 4.77 \\
All halos & $\Delta_{\rm vir}$ & 0.20 & 0.12 & 0.36 & $-0.43$ & 0.12 & 1.09 & 12.03 & 1.09 & 2.54 & 3.98 \\
\hline
Relaxed halos & $\Delta_{200c}$    & 0.20 & 0.11 & 0.33 & $-0.50$ & 0.15 & 1.04 &  9.97 & 0.68 & 1.64 & 5.59 \\
All halos & $\Delta_{200c}$    & 0.20 & 0.11 & 0.35 & $-0.47$ & 0.13 & 1.06 &  8.81 & 0.95 & 2.31 & 3.06 \\
\hline
\end{tabular*}
\vspace{1.5ex}
\begin{minipage}{\textwidth}
\textbf{Note.} The baseline CDM growth factor is given by Equation~(\ref{eq:app_Dzf_cdm}), the WDM modification factor $\mathcal{E}$ by Equation~(\ref{eq:app_EM_peak}), and the peak-concentration relation by Equation~(\ref{eq:app_c_peak_model}).
\end{minipage}
\end{table*}

As shown in Table~\ref{tab:app_peak_params}, relaxed halos consistently exhibit a higher normalization $A_{\rm p}$ and a flatter slope $\alpha_{\rm p}$ compared to the all-halo population, reflecting their systematically higher concentrations and earlier collapse epochs. 
Crucially, the cutoff coefficients $a$ determined here (last column of Table~\ref{tab:app_peak_params}) directly set the scaling of $A_{\rm hm}$ in Equation~(\ref{eq:scut}), ensuring seamless physical consistency between our individual halo model and the population peak statistics.

\bibliography{references}{}
\bibliographystyle{aasjournalv7}

\end{document}